\documentstyle[referee]{mn}
\def\sc40{OGLE-1999-BUL-19}
\newcommand{\reference}{\bibitem}
\def\mnras{MNRAS}
\def\araa{ARAA}
\def\aap{A\&A}
\def\apj{ApJ}

\def\plotone#1{\centering \leavevmode
\epsfxsize=\columnwidth \epsfbox{#1}}

\def\thetaE{\theta_{\rm E}}

\def\beq{\begin{equation}}
\def\eeq{\end{equation}}
\def\bey{\begin{eqnarray}}
\def\eey{\end{eqnarray}}
\def\kms{{\rm \,km\,s^{-1}}}

\def\chisq{\chi^2}

\def\tE{t_{\rm E}}
\def\rEt{\tilde{r}_{\rm E}}

\def\u0{u_0}

\def\Ds{D_{\rm s}}
\def\Dl{D_{\rm l}}
\def\mI0{m_{I,0}}

\def\fl{f_{\rm s}}

\def\pirel{\pi_{\rm rel}}

\def\m{~\rm{mag}}

\newcommand{\up}[1]{\ifmmode^{\rm #1}\else$^{\rm #1}$\fi}

\newcommand{\arcd}{\ifmmode^{\circ}\else$^{\circ}$\fi}
\newcommand{\arcm}{\ifmmode{'}\else$'$\fi}
\newcommand{\arcs}{\ifmmode{''}\else$''$\fi}

\input epsf
\begin{document}

\title[
Optical Gravitational Lensing Experiment. The First Multi-Peak Parallax
Event
]
{
Optical Gravitational Lensing Experiment.
\sc40: The First Multi-Peak Parallax
Event
}
\author[Smith, Mao, Wo\'zniak, Udalski et al.]
{
Martin C. Smith$^1$,
Shude Mao$^1$,
P. Wo\'zniak$^{2}$,
A. Udalski$^3$, M. Szyma\'nski$^3$,
\newauthor M. Kubiak$^4$, G. Pietrzy\'nski$^{4,3}$,
I. Soszy\'nski$^3$, K. \.Zebru\'n$^3$
\thanks{e-mail: (msmith,smao)@jb.man.ac.uk, wozniak@lanl.gov,
(udalski,msz,mk,pietrzyn,soszynsk,zebrun)@astrouw.edu.pl
}
\thanks{
Based on observations obtained with the 1.3 m Warsaw
Telescope at the Las Campanas Observatory of the Carnegie
Institution of Washington.
}
\\
\smallskip
$^{1}$ Univ. of Manchester, Jodrell Bank Observatory, Macclesfield,
Cheshire SK11 9DL, UK \\ 
$^{2}$ Los Alamos National Laboratory, MS D436, Los Alamos,
NM 87545, USA \\
$^{3}$ Warsaw University Observatory, Al. Ujazdowskie 4,
00-478 Warszawa, Poland \\
$^{4}$ Universidad de Concepci{\'o}n, Departamento de Fisica,
Casilla 160--C, Concepci{\'o}n, Chile
}
\date{Accepted ........
      Received .......;
      in original form ......}

\pubyear{2002}

\maketitle
\begin{abstract}
We describe a highly unusual microlensing event, \sc40. Unlike most
standard microlensing events, this event exhibits
multiple peaks in its light curve. The Einstein radius crossing time
for this event is approximately one year, which is unusually long. We
show that the additional peaks in the light curve can be caused by the
very small value for the relative transverse velocity of the lens
projected into the observer plane ($\tilde{v} \approx 12.5\pm 1.1
\kms$). Since this value is significantly less than the speed of the
Earth's orbit around the Sun ($v_{\oplus} \sim 30 \kms$), the motion
of the Earth induces these multiple peaks in the light curve. This
value for $\tilde{v}$ is the lowest velocity so far published and we
believe that this is the first multiple-peak parallax event ever
observed. We also found that the event can be somewhat better
fitted by a rotating binary-source model, although this is to be
expected since every parallax microlensing event can be exactly
reproduced by a suitable binary-source model. A face-on rotating
binary-lens model was also identified, but this provides a
significantly worse fit. We conclude that the most-likely cause for this
multi-peak behaviour is due to parallax microlensing rather than
microlensing by a binary source. However, this event
may be exhibiting slight binary-source signatures in addition to these
parallax-induced multiple peaks. 
With spectroscopic observations it is 
possible to test this `parallax plus binary-source' hypothesis and (in
the instance that the hypothesis turns out to be correct) to 
simultaneously fit both models and obtain a measurement of the lens
mass. Furthermore, spectroscopic observations could also supply
information regarding the lens properties, possibly providing
another avenue for determining the lens mass. We also
investigated the nature of the
blending for this event, and found that the majority of the $I$-band
blending is contributed by a source roughly
aligned with the lensed source. This implies
that most of the $I$-band blending is
caused by light from the lens or a binary
companion to the source. However, in the $V$-band, there appears to be
a second blended source 0.34 arcseconds away from the lensed source.
Hubble Space Telescope observations will be very useful for understanding
the nature of the blends. We also suggest that a radial
velocity survey of all parallax events will be very useful for further
constraining the lensing kinematics and understanding the origins of
these events and the excess of long events toward the bulge. 
\end{abstract}

\begin{keywords}
Gravitational Microlensing, Galaxy: Bulge, Galaxy: Centre,
Stars: Binaries, Galaxy: Kinematics and Dynamics

\end{keywords}

\section{Introduction}

At the time of writing, more than one thousand microlensing events
are known. In addition to the original goal
to search for the dark matter (Paczy\'nski 1986),
these events have also developed diverse applications
(see Paczy\'nski 1996 for a review).
Most of these events are well described by the standard shape
(e.g., Paczy\'nski 1986). Unfortunately, from these standard
microlensing light curves, the lens distance and mass cannot
be uniquely determined (see \S2). This degeneracy is one
of the major obstacles in studies of microlensing.

%
Fortunately, some microlensing events show deviations
from the standard shape. The parallax microlensing events
are one class of these exotic microlensing events (Gould 1992). These
events allow one to derive the projected Einstein radius (or 
equivalently, the transverse velocity) in the ecliptic plane.
This additional constraint partially lifts the lens degeneracy, but is
not enough to uniquely determine the mass of the lensing object; to do
this some other piece of information is required, such as the angular
Einstein radius. This is a rare occurrence, but a striking example of
this can be seen in An et al. (2002), where parallax signatures were
observed in a caustic-crossing binary-lens event; this completely
broke the lens-mass degeneracy and proved to be one of the first ever
measurements of the microlens mass. Alcock et al. (2001a) also
claim to have made a determination of the microlens mass for a
different event by utilising measurements of both the parallax effect
and the microlens proper motion. However, this mass determination
relies on their photometric measurement of the parallax effect, which
in this instance is very small and requires confirmation (this can be
done by obtaining a measurement of the astrometric parallax using HST,
for example). Other approaches to this degeneracy problem include
resolving the components of the lensed object, which should become
possible in the near-future with the availability of suitable optical
long baseline interferometry (see Delplancke, G\'orski \& Richichi
2001). However, in the absence of further information the
parallax effect can be combined with a model for the lens kinematics,
which allows important constraints on the lens to be drawn.
So far about 10 microlensing parallax
events have been found (Alcock et al. 1995; Mao 1999;
Smith, Wo\'zniak \& Mao 2002;
Bennett et al. 2001; Mao et al. 2002; Bond et al. 2001). Three of these
events are particularly interesting 
(Bennett et al. 2001; Mao et al. 2002) because the long Einstein radius
crossing time and the kinematics imply that the lenses are very likely
intervening black holes 
(see also Agol et al. 2002).
This is particularly exciting because these black
holes may be outside the gas layer of the disk, and hence have no
accretion signatures for detection in any
other wavelengths such as X-ray and radio. Microlensing
may be the only method to provide a complete census of the 
massive black holes in the Milky Way.

Nearly all of the published microlensing parallax events are identified
by a slight asymmetry in the light curve due to the Earth's motion
around the Sun. During the systematic search of Smith et al. (2002), which
analysed the 520 microlensing events published in Wo\'zniak et
al. (2001), one event has been uncovered that can be very well fitted
by the parallax model. This event shows a striking multi-peak signature.
Such multi-peak events have been predicted by Gould, Miralda-Escude,
\& Bahcall (1994). The purpose of this paper is to present a detailed
analysis of this event. The outline of the
paper is as follows: in Section 2 we present the observational data,
the data reduction procedure and our method to select the parallax
events from the microlensing database. In Section 3 we present the best
parallax model, while in Section 4 we explore whether \sc40 can be
fitted by alternate rotating binary-source and binary-lens
models. Section 5 discusses the various approaches that can be
utilised to provide a measurement of (or strong constraints on) the
lens mass. Finally, in Section 6, we summarize and further discuss our
results.

\section{Observations, Data Reduction and Selection Procedure}

The observations presented in this paper were carried out during the
second phase of the OGLE experiment with the 1.3\,m Warsaw telescope
at the Las Campanas Observatory, Chile. The observatory is operated
by the Carnegie Institution of Washington. The telescope was
equipped with the `first generation' camera with a SITe
2048$\times$2048 pixel CCD detector working in the drift-scan mode.
The pixel size was 24$\mu$m, giving the scale of 0.417\arcs\, per pixel.
Observations of the Galactic bulge fields were performed in the
`medium' speed reading mode with the gain 7.1 e$^{-}$ ADU$^{-1}$
and readout noise about 6.3 e$^{-}$. Details of the instrumentation
setup can be found in Udalski, Kubiak \& Szyma\'nski (1997).
The majority of the OGLE-II frames were taken in the {\it I}-band,
roughly 200-300 frames per field during observing seasons 1997--1999.
Udalski et al. (2000) gives full details of the standard OGLE
observing techniques, and the DoPhot photometry
(Schechter, Mateo \& Saha 1993) is available from the OGLE
web site\footnote{http://www.astrouw.edu.pl/\~\,ogle/ogle2/ews/ews.html}.

\sc40 was identified during a search through a catalogue of
microlensing events that had been compiled from the three year
OGLE-II bulge data. This catalogue, which was generated using the difference
image analysis technique, is available
electronically\footnote{http://www.astro.princeton.edu/\~\,wozniak/dia/lens/}
and interested readers are referred to Wo\'zniak et al. (2001) and
Wo\'zniak (2000) for further details. The aim of the search through
this catalogue was to identify potential parallax microlensing events,
and the details of this procedure can be found in Smith et al. (2002).

In fact, this event was first detected by the OGLE
Early Warning System as \sc40.
Throughout this paper we shall refer to this event as \sc40, although
in the difference image analysis catalogue of Wo\'zniak
et al. (2001) it is labelled sc40\_2895.
The position
of the star is RA=17:51:10.76, and DEC=$-$33:03:44.1 (J2000). The
Galactic coordinates are
$l=357^\circ.077,\:b=-3^\circ.147$, and the ecliptic longitude and
latitude are $268^\circ.126$ and $-9^\circ.637$, respectively.
The total {\it I}-band magnitude of the lensed star plus 
blend(s) is about $I\approx 16.07$.
The average $V-I$ colour of the composite is about $2.63 \pm 0.05$.
Fig. \ref{fig:cmd} shows the colour-magnitude diagram for the stars
within a field of view $5\arcmin \times 5\arcmin$ around \sc40. From
this figure it is clear that the star is located in the red clump region.

Initially we analysed just the three season data from 1997 to 1999
(available online\footnotemark[4]; Wo\'zniak et al. 2001). However, we
noticed that the parallax model predicts a huge spike in the 2000
season. In order to test this, we subsequently
analysed the data from this season. Reassuringly, this confirmed the 
existence of a huge spike, already seen (but unknown to the modellers)
by the OGLE Early Warning System. The four-season
data from the difference image analysis
are shown in Fig. \ref{paralc}.
In total, there are 317 data points in the light
curve. In the next section we present both the best standard
and parallax model for this unique event, while in \S4 we explore the
alternate models.

The data for this event can be found online\footnote{http://bulge.princeton.edu/\~\,ogle/ogle2/OGLE-1999-BUL-19.html}.
This site includes the full sequence of $I$-band subframes for this
event, the finding chart for the star, the DoPhot photometry, and the
full 4-season difference image analysis data that was used to model
this event.

\section{Standard and Parallax Models}

First, \sc40 is fit with the standard single microlens model.
In this model the (point) source, the lens and the observer are all
assumed to move with constant spatial velocities. The standard light curve,
$A(t)$ is given by (e.g., Paczy\'nski 1986):
\beq \label{amp}
A(t) = {u^2+2 \over u \sqrt{u^2+4}},~~
u(t) \equiv \sqrt{\u0^2 + \tau(t)^2},
\eeq
where $\u0$ is the impact parameter (in units of the Einstein radius) and 
\beq \label{tau}
\tau(t) = {t-t_0 \over \tE}, ~~ \tE = {\tilde{r}_{\rm E} \over \tilde{v}},
\eeq
with $t_0$ being the time of the closest approach (i.e., maximum
magnification), $\tilde{r}_{\rm E}$ the Einstein radius projected into the 
observer plane, $\tilde{v}$ the lens transverse velocity relative to 
the observer-source line of sight, also projected into the observer
plane, and $\tE$ the Einstein radius crossing time.
The Einstein radius projected into the observer plane is given by
\beq \label{rE}
\tilde{r}_{\rm E} = \sqrt{4 G M \Ds x \over {c^2 (1-x)}},
\eeq
where $M$ is the lens mass, $\Ds$ the distance to the source and
$x=\Dl/\Ds$ is the ratio of the distance to the lens and the distance
to the source. 
Equations (\ref{amp}-\ref{rE}) show the well-known lens degeneracy in
standard microlensing light curves. In this model the only measurable
quantity that holds any information about the lens properties is
$\tE$. This means that for a given value of $\tE$ one cannot infer
$\tilde{v}$, $M$ and $x$ uniquely, even if the source distance is
known.

The flux difference obtained from difference image analysis can be
written as
\beq \label{eq:f}
f(t) = \fl \left[A(t) - 1\right] +\Delta f,
\eeq
where $\fl$ is the baseline flux of the lensed source, and $\Delta f
\equiv f_0-f_R$ is the difference between the baseline flux ($f_0$)
and the flux of the reference image ($f_R$). All the fluxes here are in
units of 10\,ADU and can be converted into $I$-band magnitudes using the
transformation given in Wo\'zniak et al. (2001)\footnote{
$
m_I(t) = m_{I,R} - 2.5 \times {\rm log}_{10}
\frac{f(t)}{f_R},
$
where $f_R=1048.01$ and $m_{I,R}=15.86$ are the reference flux and
$I$-band magnitude, respectively.
}.
This value $f_0$ is composed of both the baseline flux of the lensed
star plus the flux from the lens and/or 
any unlensed blended star(s), if present.
Note that in general $\Delta f$ does not have to be zero or even
positive as the reference image can be brighter than the 
true baseline image ($f_R>f_0$).
For \sc40, the reference
image flux is $f_R=1048.01$ (Wo\'zniak et al. 2001). To fit
this {\it I}-band data with the standard model, we need the five
parameters given above: $\fl$, $\Delta f$ (or $f_0$), $\u0,
t_0$, and $\tE$. Best-fit parameters (and their errors) are found by
minimizing the usual $\chisq$ using the MINUIT program in the CERN
library$\footnote{http://wwwinfo.cern.ch/asd/cernlib/}$.

\begin{table}
\begin{center}
\caption{The best standard model (first row) and the best parallax
model (second row) for \sc40. The parameters are explained in \S3.
} 
\label{paratab}
\vspace{0.3cm}
\begin{tabular}{ccccccccc}
Model & $t_0$ & $\tE$ (day) & $\u0$ & $\fl$
& $\Delta f$ & $\psi$ & $\rEt$ (AU) & $\chisq$/dof \\
\hline
$S$ &
$ 1628.146\pm       0.044$ &
$    166.9\pm     1.6$ &
$    0.1075\pm    0.0014$ &
$    783\pm    11$ &
$   -124.66\pm       0.88$ &
--- & --- & 17912/312
\\
P & $ 1628.43\pm 0.21$ &
$    372.0\pm 3.3$ &
$    -0.469\pm 0.004$ &
$    706.0_{-8.3}^{+8.4}$ &
$   -188.0\pm 1.1$ &
$    3.1369_{-0.0011}^{+0.0010}$ &
$    2.684_{-0.023}^{+0.024}$ &
590.1/310
\end{tabular}
\end{center}
\end{table}

The best-fitting light curve for the standard model is shown by the
dotted line in
Fig. \ref{paralc}, and the fit parameters are presented in Table
\ref{paratab}. Clearly this model comprehensively fails to reproduce the
behaviour shown by the data. The $\chi^2$ per degree of freedom is
greater than 50, and so this model can be discounted unequivocally.

The next logical step is to attempt a fit that incorporates the
parallax effect, since the
duration of the event is particularly long: as can be seen
from the light curve, $\tE$ is in the region of a few hundred
days, during which time the Earth will have moved
substantially in its orbit around the Sun. This
invalidates the standard model's assumption that the
observer moves with a constant spatial velocity. The Earth's
centripetal acceleration induces a perturbation on the light curve, and
this can become important for events with time-scale greater than a few
months. Parallax deviations are expected to be 
especially prominent for events where the relative
transverse velocity of the system is small (i.e., comparable
to the Earth's orbital speed), since this means that the Earth's orbit has a
more significant effect on the trajectory. 
Another factor that affects the magnitude of the parallax deviations
is the size of the Einstein radius projected into the observer
plane. This radius determines the length scale on which the
magnification is calculated, and so the important quantity is the
magnitude of the Earth's motion relative to this projected Einstein
radius. Therefore, if the projected Einstein radius is significantly
greater than 1 AU then the Earth's motion will have less effect than
if the projected Einstein radius is comparable to 1 AU.

The description of the parallax effect requires two additional
parameters to represent the lens trajectory. We describe the trajectory
of the lens in the ecliptic plane, following the natural formalism
advocated by Gould (2000), and use the two additional parameters: 
$\tilde{r}_{\rm E}$, the Einstein radius
projected into the observer plane; and $\psi$,
an angle in the ecliptic plane describing the orientation of
the lens trajectory (given by the angle between the heliocentric
ecliptic $x$-axis and the normal to the trajectory). The details of this
procedure, including a figure illustrating the geometry of the situation,
can be found in Soszy\'nski et al. (2001) and will not be repeated
here (see also Alcock et al. 1995; Dominik 1998a). Once these two
additional parameters have been determined then the trajectory of the
lens in the ecliptic plane is completely determined, allowing a
calculation to be made of the separation between the lens and the
observer (i.e., the quantity that is analogous to the $\u0$ parameter
from the standard model). The light curve can then be calculated from
this separation.

The standard model's best-fit values are taken as initial guesses for
the parameters $\fl$, $\Delta f$, $\u0, t_0$, and $\tE$. However,
initial values of $\rEt$ and $\psi$ are arbitrarily chosen for a
number of combinations to search for any degeneracy in the parameter
space. The best-fitting parameters are again found by minimizing the
$\chi^2$. 
Notice that the definitions of $\u0$ and $t_0$ are slightly different
from the standard model: $\u0$ now corresponds to the closest approach
of the lens trajectory to the Sun in the ecliptic plane, and $t_0$ is
the time of this closest approach. $\u0$ is defined such that the Sun
lies on the left-hand side of the lens trajectory for positive values
of $\u0$. Unlike the standard model, these two
parameters no longer have straightforward intuitive meanings because
of geometric projections and the parallax effect. For example, the
lens trajectory's closest approach in the ecliptic plane does not, in
general, correspond to the closest approach in the lens plane and
hence does not match the peak of the light curve.

The inclusion of the parallax effect results in a dramatic reduction
in the $\chi^2$ value. The new fit has a $\chi^2$ value of 590 for 310
degrees of freedom, compared to the standard fit's value of just under
18000. This improvement can clearly be seen from the light curve
plotted in Fig. \ref{paralc}. The standard fit is unable to reproduce
the two small bumps on either side of the main peak (shown in the two
insets on the bottom panel of Fig. \ref{paralc}), resulting in a
wholly unfeasible $\chi^2$ value. However, the parallax model has no
such difficulties and both bumps are suitably fit.

The accuracy of the parallax fit is highlighted in the top panel of Fig.
\ref{paralc}, which shows the difference between the data points and
the parallax model. Any problems with the parallax fit would manifest
themselves as systematic deviations from the horizontal axis; apart
from a slight under-prediction of the flux in the second season, the
flux consistently matches the value  predicted by the parallax
model.
This $\chi^2$ value of 590 for the parallax fit corresponds to a $\chi^2$
per degree of freedom of 1.90, a formally unacceptable value. However,
from the top panel of Fig. \ref{paralc} it is clear that most of the
excess scatter occurs near the beginning of the fourth season, when
the source flux was strongly magnified. Obviously, the Poisson photon noise expressed in
counts should rise as the square root of the magnified flux. However, there are
also additional sources of noise for brighter stars in most ground
based CCD imaging. Most of these effects in OGLE-II data are accounted
for by a simple rescaling of the error bars (Fig.~3 in Wo\'zniak
2000). Near the peak of the event at $I \approx 13.8$ magnitudes, the
r.m.s. scatter is about 1\% in the present data, fully consistent
with the scatter for other comparably bright objects in this field.
The problem is that the average scaling curve for all fields may not fully
reflect the photometric uncertainty for bright stars in this part
of the BUL\_SC40 field. The error bars may be under-estimated in this region
which then result in an
artificially large $\chi^2$ value. A possible explanation might be
the larger than average extinction in this field,
contributing to lower signal to noise
in the fitted kernels and point spread functions. To ensure that these
observed systematics are not an artifact of our method of
photometric data reduction (i.e., difference image analysis), we made a
parallax fit to the regular photometry obtained with DoPhot. The
corresponding residual plot, when converted from magnitudes into counts, 
shows all the same features that have been discussed above, but with
a marginally larger overall scatter. Therefore the following analysis is
based solely on the difference image analysis data.

The model parameters for the best-fitting parallax model are presented
in Table \ref{paratab}. It can be seen from this table that all of
the parameters,
and most notably the ones describing the lens trajectory ($\u0$ and
$\psi$), are very tightly constrained, due to the fact that
the parallax deviations are especially pronounced. 
Another aspect of the light curve that
probably contributed to the tightly-constrained nature of the
parameters was the brightness of the event, since this reduces the
size of the photometric error bars. The total {\it I}-band baseline
magnitude for this event is approximately 16.07 magnitudes, and the
parallax model predicts that the lensed star contributes 82 percent of
this, i.e., there is only slight blending and the baseline magnitude
of the lensed star alone is 16.3 magnitudes. The lensed star was also
quite strongly magnified at the peak, rising to a magnification of
greater than 10 ($A_{\rm max} \approx 10.2$).

The parameters that help in determining the lens properties are:
\beq
\rEt=2.68\pm 0.23 \,{\rm AU}, ~~
\tE=372.0\pm 3.3 \,{\rm day}, ~~
\eeq
From these an expression for the lens mass can be
determined, although due to the degeneracy inherent in the
microlensing light curve this can only be expressed as a function of
the relative lens-source distance (see Soszy\'nski et al. 2001; Gould
2000),
\beq \label{mass}
M = \frac{c^2 \tilde{r}_{\rm E}^2}{4G} \left( {{1 \over \Dl}
- {1  \over \Ds} } \right)
=0.088 M_\odot \left( \frac{\tilde{r}_{\rm E}}{\rm2.68\, AU} \right)^2
\left( \frac{\pirel}{0.1 \rm mas} \right),
~~\pirel\equiv{{{\rm AU} \over \Dl} - {{\rm AU} \over \Ds} }.
\eeq
As can be seen from this equation, the lens mass depends on 
the relative lens-source parallax, $\pirel$. 
Fig. \ref{fig:cmd} suggests that the star is a red-clump star in the
bulge, so its distance is approximately 8\,kpc. If the lens is
lying half-way in-between (i.e., $D_{\rm s}\approx 8\,$kpc and $D_l \approx
4\,$kpc), then $\pirel=0.125\,{\rm mas}$, and
this would imply a lens mass of around $0.11 M_\odot$.
However, if the lens is very close to us, say $D_{\rm l} \sim 550$\,pc
(which corresponds to $\pirel > 1.6  \rm mas$),
then this lens mass can be as large as $1.4 M_\odot$.

The value of the transverse velocity projected into the observer plane
suffers from no such degeneracy and can be determined uniquely,
\beq
\tilde{v} = \frac{\tilde{r}_{\rm E}}{\tE} = 12.5\pm 1.1 \mbox{ km s}^{-1}.
\eeq
This velocity is exceptional in that it is the
lowest ever recorded for a published parallax microlensing
event. Possible causes for such a low velocity are discussed in \S6.

The exaggerated nature of the parallax signatures can be
accounted for when it is noted that the projected lens
velocity, $\tilde{v}$, is much less than the orbital speed of the
Earth ($v_\oplus \approx 30 \mbox{ km
s}^{-1}$). Because of this, as the lens trajectory passes the Earth
the change in separation is dominated by the motion
of the Earth, rather than the motion of the lens as in normal
microlensing events. The trajectory of the
lens relative to the observer-source line of sight is shown in the
top panel of Fig. \ref{paratraj},
and the highly non-linear nature can clearly be seen. During the 6
months before the point of closest approach the Earth swung away from
the lens and increased the separation, with the maximum separation
occurring at around 1450 days. This can be contrasted with the
behaviour during the 6 months directly preceding this region
(i.e., approximately $1250 < t \equiv {\rm JD} - 2450000 < 1450$
days), where the Earth completed
another half-orbit and its trajectory brought it back towards the
lens. These two regions correspond, respectively, to the unusual
declining and rising sections of the light curve which occur before the
main peak, i.e., the parallax induced `bump' at around $t=1300$ days.
Similarly, this behaviour is repeated in the 12 months following the 
point of closest approach: during the 6 months directly following the
point of closest approach the Earth moves away from the lens,
resulting in a sharp decline in flux; however, in the subsequent 6
month period the Earth begins to move back toward the lens resulting
in a sharp rise in flux. This period of one year, from approximately
$t=1650$ to $t=2000$ days, corresponds to the trough around $t=1800$
days.

The highly unusual nature of this event is highlighted when its
trajectory is compared to that of another, more typical, parallax
affected light curve. The bottom panel of Fig. \ref{paratraj} shows the
trajectory for the microlensing event sc33\_4505 which was discovered
during the parallax search through the OGLE-II database of
microlensing events toward the galactic bulge (see Smith et
al. 2002). This event had unexceptional values of $\tilde{r}_{\rm E}$
and $\tilde{v}$ (6.37 AU and $57\,\mbox{kms}^{-1}$ respectively), and
the corresponding trajectory can be seen in Fig. \ref{paratraj} to be
much closer to the standard linear approximation than
for \sc40. This is because the velocity of the lens is about twice the
size of the Earth's velocity; in the time that it takes the
lens to cross its projected Einstein diameter the Earth has completed
approximately one orbit, whereas for \sc40 the Earth completes nearly
two orbits in this time. Another factor that leads to the parallax
effect being more pronounced in \sc40 is that the value of the
projected Einstein radius, $\tilde{r}_{\rm E}$, is much smaller than
for sc33\_4505. This effect has been discussed above, and its
influence can clearly be seen in Fig. \ref{paratraj}.

Since the magnification is dependent on the magnitude of the
separation of the lens from the observer-source line of sight, a plot
of how this separation varies with time is helpful to elucidate the 
situation. This is shown in Fig. \ref{parasep}. The separation for the
\sc40 event (given in the upper-left panel) clearly varies significantly
from the hyperbolic shape of a standard microlensing light
curve (shown as a dashed line); the Earth's orbital motion induces an
annual oscillation onto this standard hyperbolic form, and from this
oscillation one can unmistakably identify the origins of the two
additional parallax bumps.
However, the situation is much less clear for the sc33\_4505 event;
the oscillations are much smoother and much smaller in magnitude. The
origins of the additional peaks and troughs in the light curve of
\sc40 can easily be identified from this figure, since they obviously
correspond to the points at which the rate of change of separation,
$\dot{u}$, is zero. For a typical
microlensing event $\dot{u}=0$ only occurs once at the point of
closest approach. However, when the projected lens velocity is so low,
as is the case in \sc40, the motion of the Earth can cause there to be
additional local points of closest approach, in addition to the global
point of closest approach.

\section{Alternate Models}

\subsection{Binary-source model}

It is necessary to test whether this unusual behaviour could be caused
by any other phenomena, such as a rotating binary source or binary lens. 
In fact, the OGLE Early Warning System initially identified this event
as a binary-source event. To address this we first fit \sc40 with a
rotating binary-source model (using a method similar to that of
Dominik 1998a; see also Griest \& Hu 1992). However, to simplify the
process, only circular orbits are considered initially. The
generalization to elliptical orbits is considered later in
\S\ref{ellip}. Both models are described in detail in the appendix.

\subsubsection{Circular orbits}
\label{bin:circ}

A total of 12 parameters are required to describe this circular rotating
binary-source model, i.e., a
further 7 parameters in addition to the 5 from the standard
model. The motion of the two sources are described using the
following parameters: the period of the orbit, $T$; the binary-source
separation, $p$, which is given in units of the Einstein radius
projected into the source plane; the flux ratio of the first source,
${\cal F}$, i.e.,
${\cal F}=\frac{f^{(1)}}{f^{(1)} + f^{(2)}}$ where $f^{(1)}$ and
$f^{(2)}$ are the fluxes from the first and second source, respectively;
and the mass fraction of the first source, ${\cal M}$, i.e.,
${\cal M}=\frac{m^{(1)}}{m^{(1)} + m^{(2)}}$ where $m^{(1)}$ and
$m^{(2)}$ are the masses of the first and second source, respectively.
As well as these parameters, a further three angles
are required: two, $\alpha$ and $\beta$, to determine the orientation of
the orbital plane; and an angle to describe the lens trajectory in the
orbital plane, $\Theta$. Once the above 7 parameters have been
combined with those from the standard model (i.e., $t_0$, $t_E$, $u_0$,
$f_{\rm s}$, $\Delta f$) then the light curve can be calculated. It should
be noted that only the flux and mass ratios (directly) enter our parametrization,
not the individual fluxes and masses of each source. A complete
description of this model, including the procedures for calculating
the light curve, is given in the appendix A1 and will not be repeated here.

As before, the best-fit parameters are found by minimizing the
$\chi^2$. However, since the description of this model requires a
large number of parameters, the $\chi^2$ surface is very
intricate. This presents a problem when attempting to find the global
minimum, and to address this issue the initial guesses for the
parameters are chosen from a large set of randomly generated values.
In this analysis the parameter space is not searched exhaustively and
although a range of fits were found with comparable $\chi^2$ values,
only the single best-fit set of parameter values is considered.

\begin{table}
\begin{center}
\caption{The best-fit parameters for the circular binary-source model.
The parameters are explained in \S4.1 and in the appendix (\S A1). The
errors have been omitted since they were found to be misleading,
owing to the complexity of the $\chisq$ surface.
}
\label{binstab}
\vspace{0.3cm}
\begin{tabular}{ccccccc}
$t_0$ & $\tE$ (day) & $\u0$ & $\fl$
& $\Delta f$ & $\alpha$ & $\beta$ \\
\hline
$    1628.8$ &
$    408$ &
$    -0.43$ &
$    609$ &
$    -191$ &
$    -1.65$ &
$    6.14$ 
\\
& & & & & &
\\
& & & & & &
\\
 $T$ (day) & $p$ 
& ${\cal F}$ & ${\cal M}$ & $\Theta$ & $\chisq$ & $\mbox{dof}$ \\
\hline
$    368$ &
$    2.0$ &
$    0.0003$ &
$    0.169$ &
$    4.60$ &
545.20 & 305
\\
\end{tabular}
\end{center}
\end{table}

The parameter values corresponding to the best-fit model are
presented in Table \ref{binstab}. The errors have been omitted since
they were found to be misleading, owing to the complexity of
the $\chisq$ surface. It is expected that
the best-fit binary-source model should, at least, provide a fit that
is comparable to the best-fit parallax model; this is because a
binary-source model can always be found that is the exact mirror-image
of the parallax model, i.e., a fit which has all of the flux
coming from one single source (${\cal F}=0\;\mbox{or}\:1$)
\footnote{
This is because in the parallax model all of the flux is {\it
received} by one single object, i.e., the Earth.
}
, an orbital period of one year, etc.
It should be noted that this is a {\it generic} property of any
parallax microlensing event where the parallax and binary-source
models are fitted independently. However, for this event the best-fit
binary-source $\chi^2$ value is 545.2 (or 1.79 per degree of freedom),
which is a distinct improvement on the best-fit parallax value of
590.1 (or 1.90 per degree of freedom).

Fig. \ref{bin} shows the light curve for this fit, and for comparison the
best-fit parallax light curve is also included.
The minute differences between the two fits are only
discernible in the top panel of Fig. \ref{bin}. This panel shows the
difference between the data points and the parallax model, along with
the predicted binary-source flux. Clearly the two fits exhibit only
slight discrepancies, with the largest digression occurring during the
early part of the fourth season, i.e., during the period where
the error bars in the data may be unreliable (see \S3).

However, a closer inspection of the best-fit parameters shows that
whilst this model may result in an improvement in the $\chi^2$ value,
it could indeed be a reproduction of the best-fit parallax model. This
deduction is made because of two parameter values in
particular: the binary period, $T$, and the flux ratio, ${\cal F}$. The
period of this binary orbit is approximately $368$ days, which is
suspiciously close to the period of the Earth's orbit. In addition,
the flux ratio is practically zero, which implies that all
of the flux is coming from just one of the sources, i.e., analogous to
the parallax model.

Further evidence can be found by analysing the other parameters; for
example, it can be shown that the orientation of the orbital plane is
almost identical to the orientation of the ecliptic plane in the
parallax model. Similar comparisons can made for the trajectory of the
lens, the radius of the orbit and the impact parameter. From this we
conclude that the best-fit binary-source model is simply a
reproduction of the best-fit parallax model. However, this does not
explain why there should be such a large improvement in $\chi^2$
between the parallax and binary-source models. This issue is dealt
with later, in \S\ref{para+bin}.

If the source is a red clump star located in the bulge (as is
indicated by Fig. \ref{fig:cmd}), then its apparent magnitude and colour
can be estimated to be $I_{\rm s} \approx 14.15$ and $(V-I)_{\rm s}
\approx 1.19$ (see \S\ref{fss}). The star's colour
is roughly consistent with theoretical predictions
for clump giants of mass $M\approx 1\,M_\odot$ with 
solar metallicity and age 5-12\,Gyr (Table 1 in Girardi \&
Salaris 2001). If we assume that the source is
7 to 8\,kpc away from the source, then its absolute magnitude 
is also consistent with the theoretical prediction (see also
Fig. 14 in Girardi \& Salaris 2001). 
If the source mass is assumed to be $M_{\rm s} \approx 1\,M_\odot$,
this can be combined with an estimate of the source distance
($\Ds \approx 8\,\rm{kpc}$) to obtain a prediction for the separation and
radial velocity of the above binary.
Using Kepler's law, the separation of the two
sources is approximately, $a=
\left[ \frac{{\rm G}}{4\pi^2}\frac{M_{\rm s}}{1-{\cal M}}T^2 \right]^{1/3}
=1.07\,\rm{AU}$, and 
the radial velocity variations should have a semi-amplitude
of around $5.3\,\rm{km s}^{-1}$ with a period 368 days. 
Since $p$ is the binary source
separation in terms of the Einstein radius projected into the source
plane, i.e., $p=a/(D_{\rm s} \theta_E)$, we can therefore 
estimate the size of the angular Einstein radius,
\beq
\label{bs:tE}
\thetaE = \frac{{a}}{p} \times \frac{1}{\Ds}
\approx 65 \mu\rm{as}.
\eeq
This value of $\thetaE$ is very small, and it may be in contradiction with
later limits on $\thetaE$ estimated from finite source size considerations
(see \S\ref{fss} and discussion).

\subsubsection{Elliptical orbits}
\label{ellip}

Since the best-fit binary-source parameters predict that the flux
ratio is almost exactly zero, we proceeded to analyse fits which have
this flux ratio set to 0, implying that all the flux comes from one of
the sources. Because of this condition there is now a degeneracy
between $p$ and ${\cal M}$, and only the product of these two
parameters is physically meaningful ($p\,{\cal M}$ is now the
orbital semi-major axis for the luminous source). We choose to set
${\cal M}=1$, which means that $p$ now corresponds to the orbital
semi-major axis for the luminous source.

As there are now two fewer parameters, it becomes more feasible to fit
for elliptical orbits. This introduces two additional parameters: the
eccentricity, $e$; and the phase of the orbit, $\xi_0$. The details of
this model can be found in the appendix (\S\ref{app:elliptical}). If
this model is to reproduce a mirror-image of the parallax model (as
discussed above), then we would expect the eccentricity, $e$, to be
close to zero, resulting in a near-circular orbit. This is because,
unlike the Earth's orbit (which has $e_{\oplus} \sim 0.017$), it is
rare for binary stars to have nearly circular orbits (Duquennoy \&
Mayor 1991).

As with the circular binary-source model, the best fit parameters are
found by minimizing the $\chisq$. A number of initial guesses were
chosen as input parameters, and the single best-fit set of parameters
are presented in Table \ref{ellipse}.

\begin{table}
\begin{center}
\caption{The best-fit parameters for the elliptical binary-source
model, with ${\cal M}$ fixed at 1 and ${\cal F}$ fixed at 0.
The parameters are explained in \S4.1 and in the appendix (\S A1 and
\S A2). The errors have been omitted since they were found to be
misleading, owing to the complexity of the $\chisq$ surface.
}
\label{ellipse}
\vspace{0.3cm}
\begin{tabular}{cccccccc}
$t_0$ & $\tE$ (day) & $\u0$ & $\fl$
& $\Delta f$ & $\alpha$ & $\beta$ &  $T$ (day) \\
\hline
$ 1627.8$ &
$ 428$ &
$ -0.41$ &
$ 557$ &
$ -193$ &
$ -4.35$ &
$ 0.12$ &
$ 367$
\\
& & & & & &
\\
& & & & & &
\\
$ e $ & $p$ & $\xi_0$ 
& ${\cal F}$ (fixed) & ${\cal M}$ (fixed) & $\Theta$ & $\chisq$ & $\mbox{dof}$ \\
\hline
$ 0.010$ &
$ 0.32912$ &
$ 3.56$ &
$ 0$ &
$ 1$ &
$ -1.24$ &
544.84 & 305
\\
\end{tabular}
\end{center}
\end{table}

This best-fit provides a slight improvement in the $\chisq$ value
compared to the circular binary-source model. However, as can be seen
from Table \ref{ellipse}, the eccentricity is very close to Earth's
value, and the orbital period is almost exactly one year. Therefore we
conclude that this result strengthens the argument that the
binary-source model is simply reproducing a mirror-image of the
parallax model.

\subsection{Binary-lens model}

In the previous subsections, we have fitted the light curve of \sc40
with a parallax model and a rotating binary-source model. In this
section, we will explore whether \sc40 can be fitted by a rotating
binary-lens model.

While the light curves produced by the binary-source or the
parallax models are always smooth, those produced by
binary lenses may contain sharp rises and falls due
to caustic crossings (e.g., Mao \& Paczy\'nski 1991). As a result of these
new features, the $\chi^2$ surface in the multi-dimensional space is
even more complex, and it is often difficult to locate the global
minimum. Furthermore, for weak binaries and ill-sampled light curves,
the solutions are known to be degenerate (Mao \& di Stefano 1995; Dominik
1999).  For \sc40, the multi-peak behavior is
reminiscent of periodic rotations; we have therefore implemented a simple
version of rotating binaries (Dominik 1998a) where we assume the binary orbit is
face-on, i.e., the orbital plane is perpendicular to the line of sight.
Other than the five parameters in the standard model,
we have six more parameters that describe the binary
and the source trajectory: the binary lens major axis, $p$; the
angle, $\phi$, between the normal to the source trajectory and the
line connecting the lenses at the time of the perihelion; 
the period of the binary orbit, $T$; the eccentricity, $\epsilon$; the
time of the perihelion, $t_{\rm Peri}$; and the mass ratio ${\cal
M}=\frac{m^{(1)}}{m^{(1)} + m^{(2)}}$ where $m^{(1)}$ and $m^{(2)}$
are the masses of the first and second lenses respectively. Note that
$t_{\rm E}$ corresponds to the Einstein radius crossing time for the
total mass. In total,
we have 11 parameters even for this simple face-on rotating binary model.

We have searched for the best fit face-on rotating binary starting
from a number of initial guesses (although no exhaustive
searches were performed).  Fig. \ref{bin} shows the best fit
which was found. The total $\chi^2$ is 1522.4. The parameters are given in
Table \ref{binltab}.

\begin{table}
\begin{center}
\caption{The best-fit parameters for the face-on rotating binary-lens
model. The parameters are explained in \S4.2. The errors have been
omitted due to the complexity of the $\chisq$ surface.
}
\label{binltab}
\vspace{0.3cm}
\begin{tabular}{ccccccc}
$t_0$ & $\tE$ (day) & $\u0$ & $\fl$
& $\Delta f$ & ${\cal M}$ & $p$ \\
\hline
$    1536.1$ &
$    230.2$& 
$    -0.536$ &
$    1048.0$ &
$    -168.4$&
$    0.027$&
$    0.39$ 
\\
& & & & & &
\\
& & & & & &
\\
$\phi$ & $T$ (day) & $\epsilon$ & $t_{\rm Peri}$ & & $\chisq$ & $\mbox{dof}$
\\
\hline
$    1.74$ &
$    463.3$ &
$    0.26$& 
$    1368.5$ &
$    $ &
1522.4 & 306
\\
\end{tabular}
\end{center}
\end{table}
%

The $\chi^2$ is much worse
than the best parallax model and binary-source model. Much of
the $\chi^2$ actually arises from the failure of the
binary-lens model to fit the first and second season data. This
binary-lens model seems to resemble the observed shape of the light
curve, in particular the shape around the peak. However, it does not
match the data points for the first three seasons, including the
baseline. Note that in this model the lensed source contributes all of
the light.

Clearly this fit is not acceptable. It is possible, however, that an
improved fit may be found from a more exhaustive search, especially
when non-face-on orbits and blending are also considered.

\section{Constraints on the Lensing Configuration}

\subsection{Combined binary-source model incorporating parallax effect}
\label{para+bin}

As can be seen from \S4, the rotating binary-source model provides the
best fit for \sc40. However, it was shown that this model is almost
certainly a mirror-image of the best-fit parallax model, and hence not
a genuine physical solution. On the other hand, the improvement in
$\chisq$ is significant; if the error bars are rescaled so that the
$\chisq$ per degree of freedom is 1.0 for the best-fit parallax model,
then this corresponds to $\Delta\chisq = 23.7$ for an additional 5
degrees of freedom. Obviously this is a significant improvement, and
the probability of such an improvement occurring by chance is much
less than 1 percent. 

This may be accounted for by the additional parameters in
the binary-source model; an improved fit can result from a fine-tuning
of these additional parameters which are fixed in the parallax model,
e.g., orbital period or the orientation of the orbital plane. Another
possible explanation for this could be that there are systematic
errors in the data, correlated on long time-scales, which are better
fit by this rotating binary-source model.

However, a more-likely possibility is that this event is
simultaneously exhibiting both parallax
and rotating binary-source behaviour. This conclusion can be justified
when one considers that the duration of the event is significantly
greater than one year, and therefore there must be some parallax
effect on the light curve. Hence the simple rotating binary-source
fit in \S4, which did not incorporate the parallax motion of the Earth,
is obviously deficient.

Fitting the data with both the parallax and binary-source models
simultaneously would enable constraints to be put on the lens
mass. With follow-up spectroscopic observations, a rotating
binary-source fit can provide a measurement of, or at least strong
constraints on the Einstein radius projected into the source plane,
$D_{\rm s}\,\thetaE$ (Han \& Gould 1997). When combined with
the value of $\rEt$ obtained from the parallax effect, this allows one
to measure or constrain $D_{\rm s}\,M$, where $M$ is the lens
mass. Since $D_{\rm s}$ is known to be approximately 8\,kpc for \sc40, this
would at least provide strong constraints on the lens mass for this
event, and possibly a direct measurement. 

However, the task of fitting the data with both parallax and rotating
binary-source models simultaneously is not an easy one. This situation
would be helped if the period, phase and eccentricity of the binary
orbit could be determined. This can be done through radial velocity
measurements of the source, and once these are obtained then they could
be incorporated into the fitting procedure to significantly reduce the
number of free parameters. Since the Earth's orbit is known, this
means that incorporating the binary-source motion into the parallax
model should be feasible, allowing (the aforementioned) tight
constraints to be put on the lens mass.

So far, very few binary-source microlensing events have been
detected (Alcock et al. 2001b, Becker 2000). This is contrary to
expectation since many stars in the galaxy are in binaries (Griest \&
Hu 1992, Dominik 1998b, Han \& Jeong 1998). In particular, the
progenitor of the source for \sc40 was probably a G-type star on the
main sequence with a mass $\approx 1\,M_\odot$ (e.g., Table 15.8 from
Cox 2000), and it is known that the
majority of G-type stars exist in binaries. One explanation
for the lack of observed binary-source events is that to be detectable
they must have both an impact parameter which is comparable to the
binary separation, and also an orbital period which is comparable to
the event's duration, $\tE$. Typically, $\tE$ is less than a month,
and there are very few binary systems which have such short
periods. However, event \sc40 has a duration of $\tE \approx 370$ days,
which could feasibly be comparable to the orbital period of a binary
source. Even if the binary period is much greater than 370 days, deviations 
from the parallax-only fit could still be detectable in this light
curve since the microlensing amplification lasts for well over six
years in total (as
can be seen from Fig. \ref{fig:pred}). In addition, the source
for \sc40 is particularly bright and the parallax fit is well
constrained. These two factors should increase the possibility of
detecting slight deviations from the predicted parallax-only fit, even
if the period is greater than one year. If such deviations are not
apparent in the data already obtained, then they may still be observed
in the future light curve, especially with improved photometric
accuracy.

As well as the above approaches, the astrometric signature may be able
to detect the presence of a binary source. The astrometric
microlensing signature for an event lasts much longer than its
photometric counterpart (e.g., Hosokawa et al. 1993;
H${\rm \o}$g, Novikov \& Polnarev 1995; Miyamoto \& Yoshi 1995; Walker
1995; Paczy\'nski 1998), and could therefore be used to make future
observations. As was shown in Han (2001), the astrometric signature
for a binary-source microlensing event differs from that of a
single-source event, particularly when the binary-source separation is
small enough to be detectable in the photometric light curve, as may be
the case with \sc40. Therefore, if the parallax and binary-source
signatures can be dis-entangled, then this could provide another
method of determining the lens mass. However, even if the
binary-source signature is not detectable in the astrometric
observations, it is still possible to measure $\thetaE$ 
using a single-source fit to the astrometric data and hence
constrain the lens mass.

\subsection{The nature of the blend(s)}
\label{blend}

In order to select variable sources, such as microlensing events,
the difference image analysis technique automatically subtracts
out non-varying sources. However, information on the (non-varying)
blended light is also encoded in the images and can be important
for understanding the lensing geometry. Gould \& An (2002) showed that
using a proper linear weighting of the images, one can form an image
that is free of the lensed source. In their method, the weighting
of the images uses the magnification as a function of time
and the frames are convolved to the worst seeing among all frames
in the stack. In our case, we took the magnification history from
the best parallax model and constructed corresponding images of
the lensed source and separately the blended light from neighbouring
objects.

For the present application, we are particularly interested in comparing
the colours and astrometry of the blend(s) compared with those of the lensed
source. If the blended light and the lensed source are not co-aligned, then
the star which is causing this blending cannot have any influence on the
microlensing event. However, if the blended light is co-aligned with
the lensed source, then there are two very interesting
possibilities. One is that the light is from the lens itself
\footnote{For microlensing to occur, the lens and source must be
co-aligned to within a milli-arcsecond, which cannot be resolved from
the ground except using interferometry (see Delplancke, G\'orski \&
Richichi 2001).} and the other is that the light is from a (close)
binary companion source.

We applied the Gould \& An (2002) method to \sc40. We first
analysed the $I$-band images. A reference image was created by stacking
the $\sim 80$ best-seeing images. This exercise was somewhat
complicated by twelve bad columns close to the lensed star.
Fig. \ref{fig:blend} shows an $I$-band image of the lensed source
(top left) and an $I$-band image of the blend(s), i.e., with the
lensed star subtracted
(top right). The $I$-band position of the lensed star is indicated  by
a cross in all images.

From the figure it is clear that in the $I$-band the lensed star and the blend
are almost aligned. We used the DoPhot program to obtain the photometry
and astrometry of the lensed star and the blend. We found that the lensed
source and the blend are co-aligned to within 0.095 pixels (one pixel
corresponds to about 0.417 arcseconds for the OGLE-II CCD camera).
This offset is within the uncertainty of the OGLE-II astrometry,
$\sim 0.05-0.1$ pixels. The blend contributes about 24\% of the total flux,
which is comparable to but smaller than
the value predicted by the parallax model
($\sim 18\%$, see \S3). However, we do not regard
this discrepancy as very significant
given the fact that the image with the lensed source removed
is still very crowded. Also, the fact that there may be two unresolved
blends (see below) may have noticeably affected the DoPhot 
photometry. Additional possible uncertainties may arise
from correlated noise introduced by convolutions.

Next, we applied the same procedure to the $V$-band images. Most of
frames for OGLE-II were taken in the $I$-band, so only 9 $V$-band images
are available. We chose 5 frames as the other four have substantially worse
seeings. The selected $V$-band frames had slightly  better seeing than the
typical $I$-band images. Fig. \ref{fig:blend} shows the $V$-band
images of the lensed source (bottom left) and the blended light
(bottom right). In this case, the blended light shows a clear
offset. Quantitatively, the blend is about 0.35 pixels to the east,
and 0.75 pixels north of the lensed source; it contributes
about 27\% of the total flux in the $V$-band. This astrometric offset is
highly significant statistically. So while in the $I$-band the blended
light seems to be aligned with the lensed source, in the $V$-band the
blended light seems to show a substantial offset. To verify this, we
examined the centroid position of the composite as a function of
magnification. The centroid is expected to shift toward the lensed
source as its magnification increases (for an example see
Alard, Mao \& Guibert
1994). This trend is clearly seen in the $V$-band data, whilst it is absent
in the $I$-band within the astrometric uncertainties. We note that
the combined colour of the blend is similar to the composite as
the light fractions contributed by the blends are similar in the $V$-band
and in the $I$-band, a fact that we will use in \S5.3.

The most straightforward interpretation is that there are actually two blended
sources within the seeing disk of the lensed source. One is aligned with the
lensed source, and is bright in the $I$-band but faint in the
$V$-band, while the other is offset from the lensed source and is
bright in the $V$-band but faint in the $I$-band. The $V$-band image
actually shows a faint extension, or a wing, of the blend image
towards the position of the lensed source, supporting our conclusion
about the presence of a second blended source, roughly co-aligned with
the lensed source. The spatial resolution from the ground is not
sufficient to accurately determine the colours and positions of these
two blended sources. HST and ground-based spectroscopy may be
important for understanding the nature of the blending; we return to
this in the discussion.

If it is assumed that this $I$-band blend is indeed caused by the lens,
then it is possible to estimate the properties of the lens using
a mass-luminosity relationship. This assumption would give a lens
brightness of $I_{\rm l} \approx 17.9$ mag, and from the range of inferred
lens masses one can deduce that this would have to be main-sequence star.
Therefore, by comparing this luminosity (after extinction has been
taken into account)
with a mass-luminosity relationship for main-sequence stars (e.g. Cox
2000), it can be shown that the lens would probably be an early M-type or
a late K-type dwarf with a mass of around $0.6~M_{\odot}$ lying
approximately 1.3\,kpc from us. This would imply $\thetaE \sim 1.7$ mas,
which is vastly different from the value calculated from the `pure'
binary-source model in \S\ref{bin:circ} ($\thetaE \approx 65 \mu\rm{as}$).

\subsection{Finite source size}
\label{fss}

The parallax analysis in \S3 assumes that the source can be regarded as
point-like. However, as was shown in the parallax fit of \S3, the peak
magnification for this event is approximately 10, and at such large
magnifications it
becomes necessary to consider the finite size of the source. To do
this, the event is fit with a parallax model which incorporates the
finite source-size effect (see, for example, Gould 1994; Nemiroff \&
Wickramasinghe 1994; Witt \& Mao 1994). This involves an additional
parameter, $\rho_\star$, which is the size of the source in units of
the lens' angular Einstein radius.

This effect should be most prominent around the peak of the light
curve. However, as can be seen from Fig. \ref{paralc}, the peak is
reasonably well fit by the point-source parallax model. No significant
improvement was found when this finite source-size effect was
incorporated into our model; at the $2\sigma$ confidence level the following
constraint was found for $\rho_\star$,
\beq \label{rho}
\rho_\star < 0.119.
\eeq

\begin{figure}
\plotone{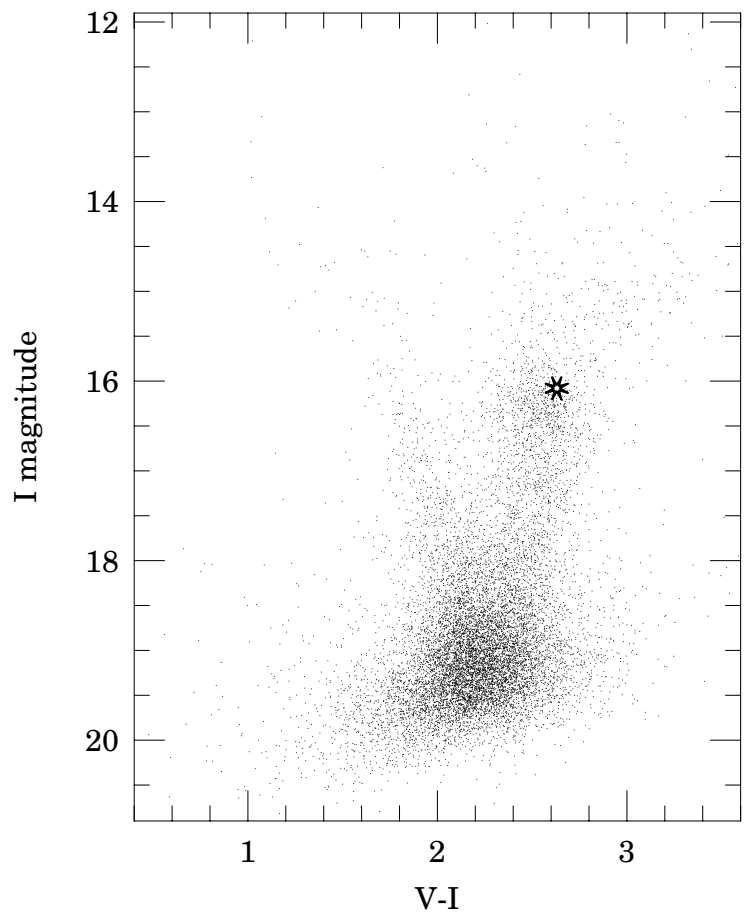}
\caption{The colour-magnitude diagram of the $5\arcmin \times 5\arcmin$
field around \sc40. The position of \sc40 is marked with an asterisk.
The star is located in the red-clump region.}
\label{fig:cmd}
\end{figure}

\begin{figure}
\plotone{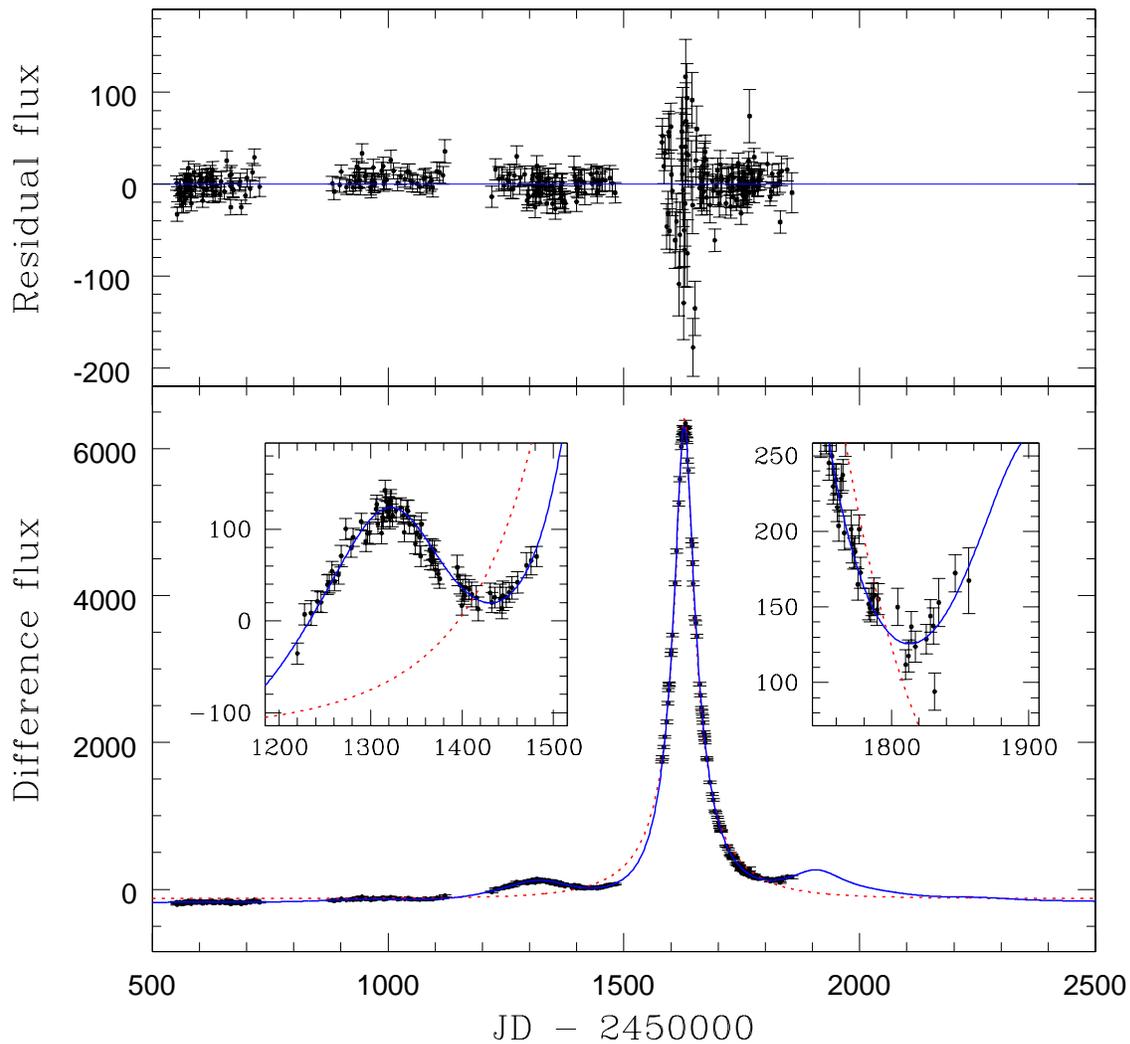}
\caption{
The {\it I}-band light curve for \sc40 from difference image
analysis, with flux given in units of 10 ADU, which
can be converted into magnitudes using the transformation given in \S3.
 The dotted and solid
lines are for the best standard and parallax fits, respectively.
The insets detail the two unusual bumps in this light curve, clearly
showing how successful the parallax model is in fitting these features.
The top panel shows the residual flux (the observed data points
subtracted by the parallax model).
The fact that there are no significant systematic deviations from the
horizontal axis indicate the goodness of the parallax fit.
The large scatter in the early part of the 4th season of data is
explained in \S3 of the text.}
\label{paralc}
\end{figure}

\begin{figure} 
\plotone{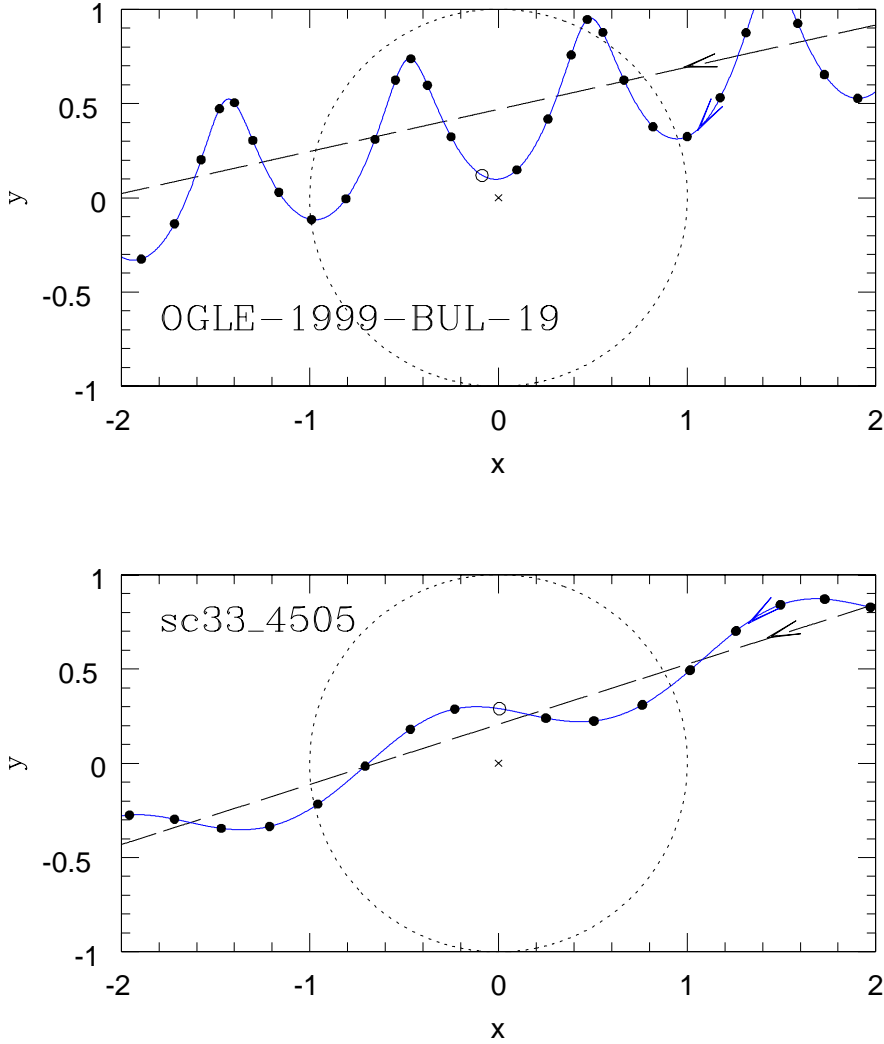}
\caption{
The trajectory of the lens (given by the best-fitting parallax model)
in the observer plane relative to the observer-source line-of-sight,
i.e., the location of the lens with respect to the Earth (denoted by
the small cross). The top
panel depicts the trajectory for \sc40, whilst the bottom panel shows
a more typical parallax event, sc33\_4505 (see \S3 and Smith et al. 2002).
This separation between the lens and the Earth corresponds to the $u$
value given in eq. (\ref{amp}) and determines the magnification for the
event. The straight dashed line represents the equivalent trajectory
without accounting for the orbital motion of the Earth (i.e., a
standard `constant-velocity' light curve). It should be noted that
this does not necessarily correspond to the best-fit standard light
curve given in Fig. \ref{paralc} and Table \ref{paratab}.
The solid dots indicate the lens position at 50 day intervals, and the
open dots correspond to the approximate peak times, i.e., $t=1650$ and
$t=650$ days for \sc40 and sc33\_4505, respectively. The large dotted
circle represents the size of the event's Einstein radius, and the
axes are given in units of this Einstein radius. For both events the
lens traverses from right to left, as depicted by the arrows. This
figure demonstrates that the trajectory in the case of event \sc40 is
highly non-linear, leading to the exceptionally dramatic parallax
signature. Comparing this to a more typical parallax microlensing
event, sc33\_4505, highlights the spectacular nature of the trajectory
for \sc40.}
\label{paratraj}
\end{figure}

\begin{figure}
\plotone{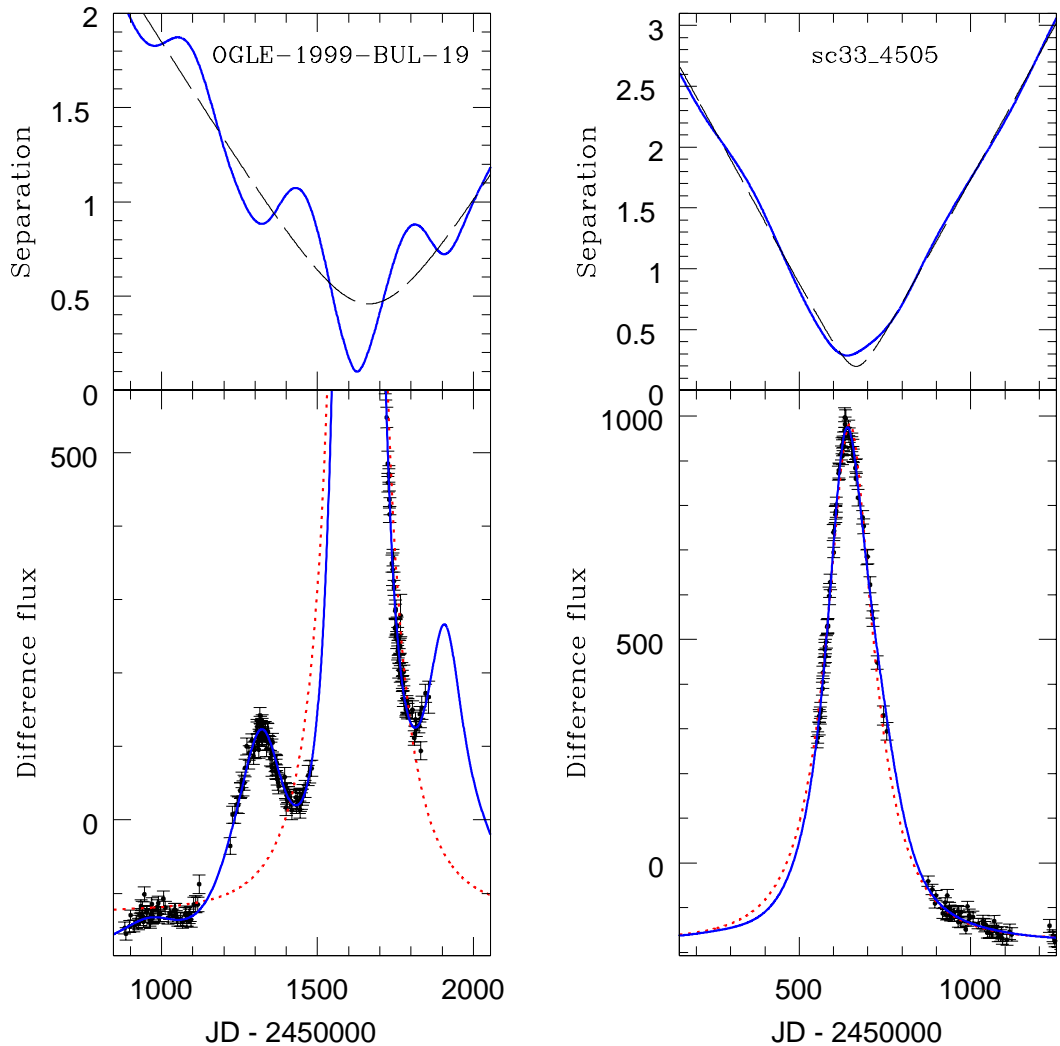}
\caption{
The magnitude of the separation (given by the best-fitting parallax
model) between the lens and the observer-source line-of-sight (top
panel), and how this relates to the light curve (bottom panel), for
\sc40 (left) and a more typical parallax microlensing event sc33\_4505
(right; see \S3 and Smith et al. 2002).
In the top panels the separation determined from the parallax fit is
given as a solid line, and the dashed line represents how this
separation would look without the orbital motion of the Earth (i.e., a
standard `constant-velocity' light curve). The separation, which is in
units of the Einstein radius, corresponds to the $u$ value given in
eq. (\ref{amp}) and determines the magnification for the event.
The lower panels show the {\it I}-band light curve for each event,
with the dotted and solid lines corresponding to the best-fit
standard and parallax models, respectively.
This figure demonstrates the highly non-linear nature of the trajectory
for event \sc40, as can be seen from the significant deviations in
separation between the parallax trajectory and the standard-type
trajectory. These deviations are clearly much more prominent in the
separation plot of \sc40, compared to sc33\_4505. From this plot one
can easily identify the origins of the irregular bumps in the light curve
of event \sc40, which correspond to the separation minima in the upper
panel. Note that the dashed line in the top panel does not necessarily
correspond to the best-fitting standard model given in the lower panel
- this is included solely to give an indication of the deviation of
the parallax separation from a linear trajectory.}
\label{parasep}
\end{figure}

\begin{figure}
\plotone{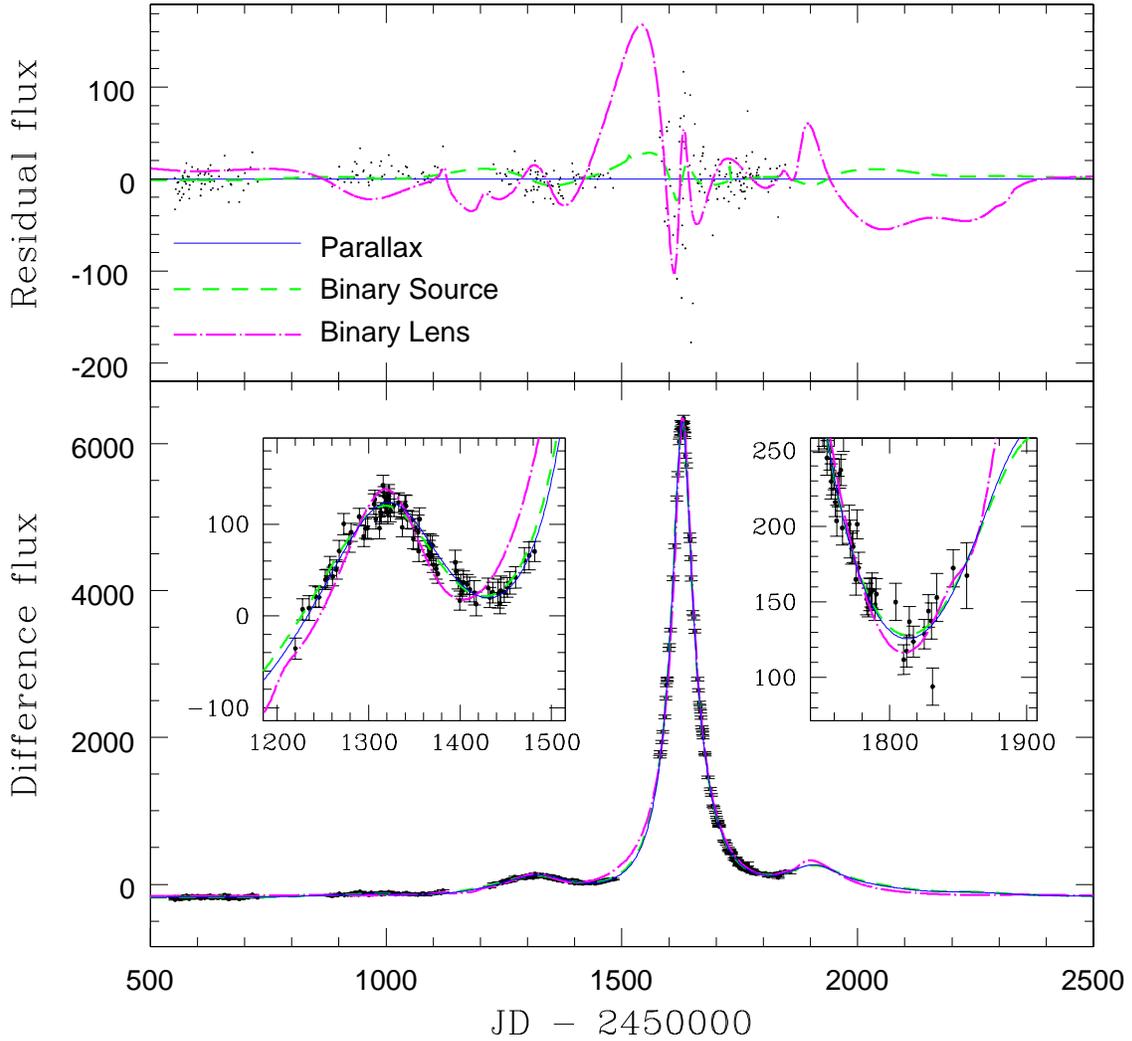}
\caption{
A diagram to compare the two best-fit binary models with the
best-fit parallax model. The bottom panel shows the usual {\it I}-band
light curve, with the best parallax model given by the solid
line, the best rotating circular binary-source model given by the dashed
line, and the best face-on rotating binary-lens model given by the
dot-dashed line. The insets detail the two unusual bumps in this light
curve. Notice that
the binary-lens model does not match the first two season data
well. The top panel shows the residual flux (the observed data points
subtracted by the parallax model). This figure shows how closely the
binary-source light curve matches the parallax light curve, with the
slight differences only becoming apparent in the residual plot.}
\label{bin}
\end{figure}

\begin{figure}
\plotone{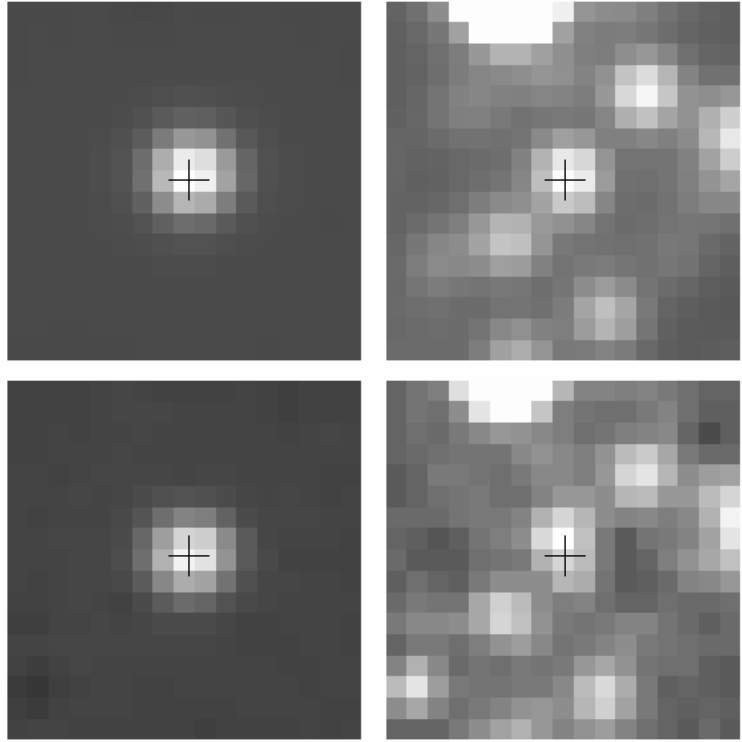}
\caption{
$I$-band (top) and $V$-band (bottom) images for the lensed source
(left-panel) and the blended light (right-panel). The crosses indicate
the position of the lensed source in each panel. Each pixel in the
panels corresponds to 0.417\arcsec. East is to the right and
north is up.
}
\label{fig:blend}
\end{figure}

\begin{figure}
\plotone{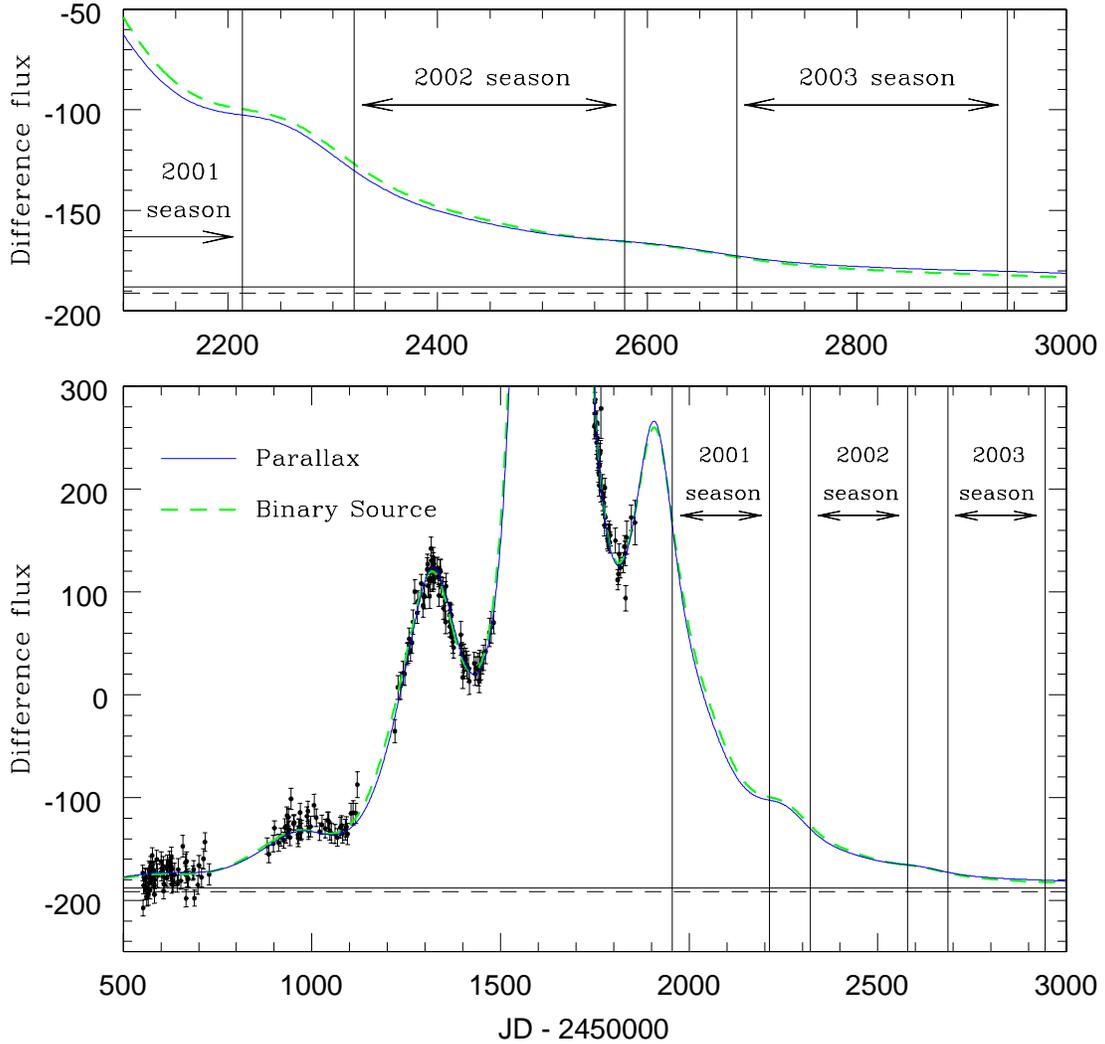}
\caption{
The {\it I}-band light curve for \sc40 showing the predicted
behaviour for the present and future observing seasons.
The solid and dashed lines
correspond to the best-fit parallax and circular binary-source models,
respectively, and each model's predicted baseline is also given. The
approximate Galactic bulge observing seasons are denoted by
vertical lines. Unfortunately, as can be seen from this figure, the final
additional peak at $t \approx 1900$ occurred during the break between
observing seasons. In addition, it should be noted that there were no OGLE
observations during the 2001 season, although data has been gathered
by the PLANET collaboration for the latter part of this season (see \S6).
The top panel details the predicted flux, and from this it is clear
that the differences between the binary-source fit and the parallax
fit are unlikely to be resolved. However, from this prediction one
should be able to test whether the parallax and binary-source models
are feasible or not, since they both predict a significant drop in
flux between the end of the 2001 season and the beginning of the 2002
season.}
\label{fig:pred}
\end{figure}

When this is combined with another independent determination of the angular
source size, a corresponding constraint can be determined for the lens
mass. As has been shown in previous work (e.g., Albrow et al. 2000),
the angular size of the source can be determined from its de-reddened
colour and magnitude. The colour-magnitude diagram for this event
(Fig. \ref{fig:cmd}), shows that the centre of the red clump
region for this field is given by $I_{\rm{cl, obs}} \approx 16.25~\rm{mag}$,
$(V-I)_{\rm{cl, obs}} \approx 2.55~\rm{mag}$. From Paczy\'nski et
al. (1999), the intrinsic de-reddened colour of the centre of the red
clump region for Baade's window is $(V-I)_{\rm{cl, 0}} =
1.114~\rm{mag}$. Therefore the reddening for \sc40's field is given
by,
\beq \label{red}
E(V-I)_{\rm{cl}} = (V-I)_{\rm{cl, obs}} - (V-I)_{\rm{cl, 0}} \approx
1.44~\rm{mag}.
\eeq
The extinction is given by (Stanek 1996),
\beq \label{ext}
A_{I,\rm{cl}}=1.49 \times E(V-I)_{\rm{cl}} \approx 2.14\m.
\eeq
The parallax model predicts that the $I$-band baseline magnitude of
the source is $I_{\rm s}=16.29\m$, and the average $(V-I)_{\rm s}$ colour of the composite
is $2.63\pm0.05\m$. From the colour-magnitude diagram for this field
(Fig. \ref{fig:cmd}), one would conclude that the source is most
likely to be located in the red clump region and hence undergo the
same reddening and extinction as calculated in eq. (\ref{red}) \&
(\ref{ext}). This would imply that the source has an intrinsic
brightness of $I_{\rm s, 0} \approx 14.15\m$ and an intrinsic colour of
$(V-I)_{\rm s, 0} \approx 1.19\m$ where we have used the fact that
the source has roughly the same colour as the composite (see \S5.2).

This colour can then be converted from $(V-I)_{\rm{s, 0}} \approx
1.19\m$ into $(V-K)_{\rm s, 0}\approx2.74\m$ (using, for example, Table III
of Bessell \& Brett 1988). Once this has been done, a value for
$\theta_\star$, the angular size of the source, can be calculated from
the following empirical surface brightness-colour relation (eq. 4
from Albrow et al. 2000),
\begin{eqnarray}
\label{tstar}
\rm{log}~(2 \theta_\ast) + \frac{V_{\rm s, 0}}{5} &=& 1.2885 \pm 0.0063
+ (0.2226 \pm 0.0133) \times [(V-K)_{\rm s, 0}-2.823]\\
\Rightarrow ~~ \theta_\ast &\approx& 7.95~\mu\rm{as}.
\end{eqnarray}

This value for $\theta_\star$ can be combined with the above
constraint on $\rho_\star$ (eq. \ref{rho}) 
to determine the lens' angular Einstein
radius,
\beq
\label{eq:thetaE}
\theta_E=\frac{\theta_\ast}{\rho_\ast} \ga~ 66.8~\mu\rm{as}.
\eeq
and hence the following constraints on the lens mass and distance
parameter $\pirel$ can be determined,
\beq
M_{\rm l} = \frac{c^2}{4G}\rEt\thetaE 
= 1.228\times10^{-4} \left( \frac{\rEt}{\rm{au}} \right) 
\left( \frac{\thetaE}{\mu\rm{as}} \right)
\ga 0.022 M_\odot,
\eeq
and,
\beq
\pirel\equiv\left(\frac{\rm au}{D_{\rm l}}-\frac{\rm au}{D_{\rm s}}\right)
=\frac{\thetaE}{\rEt/\rm{au}}\ga0.025 \rm{mas}.
\eeq
Therefore, if the source is a red clump star and it is located in the
bulge at a distance of $D_{\rm s}=8\,\rm{kpc}$, then this implies that
$D_{\rm l}~\la~6.7\,\rm{\rm kpc}$.

These constraints are not particularly severe, and this limit on $\thetaE$
is easily in agreement with the parallax value calculated in the
previous section ($\thetaE \sim 1.7$mas; \S\ref{blend}). However, the
$\thetaE$ estimate made from the binary-source model in
\S\ref{bin:circ} may contradict this limit, even though it has been
acknowledged that this constraint is not very severe. There it was
shown that $\thetaE \approx 65 \mu\rm{as}$, which would further
suggest that the `pure' binary-source model may not be a true physical
solution.

\section{Summary and Discussion}

In this paper we have shown that \sc40 is a unique microlensing event
which exhibits multi-peak behavior. The event was first identified in the
OGLE-II early-warning system. The four season data from 1996-2000 can
be reasonably fitted by a parallax model. We have also shown
that the event can be fitted by a rotating binary-source model with a
somewhat better $\chi^2$. Attempts were also made to fit the data with
a face-on rotating binary-lens model; the derived $\chi^2$ is substantially
worse, although we can not exclude the possibility that a better
binary-lens model can be found from a more exhaustive search.

It is unlikely that the predicted difference in flux between the
parallax model and the binary-source model will be discernible (see
Fig. \ref{fig:pred}).
Even though photometric observations may be unable to discriminate
between these two models, such observations will be able to test
whether these models are feasible or not, since they both
predict a significant drop in flux (approximately 0.03
magnitudes) between the end of the 2001 season and the beginning of
the 2002 season. Unfortunately, the OGLE data only covers the
1996-2000 seasons, but further data has been obtained by
the PLANET collaboration.

However, even though the rotating binary-source model provided a
better fit than the parallax model, it was shown in \S\ref{bin:circ}
that the improvement in $\chi^2$ between the two models may not mean
that the binary-source model is a more-likely solution than the
parallax model.
In fact, the parameters for this best-fit binary-source model
indicate that it is probably reproducing a mirror-image of the
parallax model. This is a generic degeneracy which exists when the
parallax and rotating binary-source models are fit independently.
We believe that the most-likely explanation for the improvement in
$\chisq$ between the parallax and binary-source models is that the
event is exhibiting both parallax and binary-source behaviour
simultaneously, i.e., the major variability is caused by parallax,
with a slight additional variation on top of this which is caused by
some binary rotation in the source (see \S\ref{para+bin}). This would
be very difficult to fit; however, if the orbital parameters of the
binary source could be constrained by spectroscopic observations, then
such a fit may be possible. Simultaneously fitting the parallax and
binary-source models would enable strong constraints to be put on the
lens mass, and may even provide a direct measurement of it.
These spectroscopic observations may
also confirm the `pure' binary-source model, in the extremely unlikely
case that this is proved to be correct. It was shown in
\S\ref{bin:circ} that the amplitude of the radial velocity variations
predicted by this binary-source model should be approximately
$5\rm{km s}^{-1}$ with a period of one year, and therefore this
prediction can easily be discounted by a couple of measurements with
$1\rm{km s}^{-1}$ precision. However, as was noted in \S\ref{fss},
these predictions from the `pure' binary-source model may violate the
limit placed on $\thetaE$ by finite source size considerations,
providing further evidence that this `pure' binary-source model may
not be a true physical solution.

One striking prediction of the parallax model is the projected
velocity $\tilde{v}$, which is the lowest seen in any lens candidate.
Smith et al. (2002) pointed out that such low-transverse 
velocities may be caused by a disk-disk lensing event. However, this
does not appear to be the case for \sc40, as the source seems to be a
red-clump star in the bulge (see Fig. \ref{fig:cmd}; see also Bennett
et al. 2001 for other black hole candidates). Usually,
a low transverse velocity is produced when the lens and source
move more or less radially (Mao \& Paczy\'nski 1996). 
Therefore it will be very important to obtain the radial
velocities of \sc40. A spectroscopic survey of all parallax events,
including \sc40, will be a worthwhile effort since radial velocities
will provide further information on the source kinematics. Such a
survey will also shed light on the related (unsolved) problem of the
nature of the excess of long events, which was first noticed by Han \&
Gould (1996).

We have investigated the photometry and astrometry of the blended sources.
The $V$-band and $I$-band data together imply that there are two
blended sources within the seeing disk of the lensed source, one is
aligned with the lensed source while the other is mis-aligned by about
0.83 pixels ($\sim 0.37\arcsec$). This mis-aligned blend could be an
unrelated star or, if the lens is close enough to the observer, this could
be a wide-separation companion to the lens (although if the lens is at
a distance of 1.3kpc, as was suggested in \S\ref{blend}, then the
separation would be around 400 au, which would be too great a
separation for this companion to have any influence on the microlensing
light curve). However, the OGLE-II data are not
sufficient to decipher these two components. The Advanced Camera for
Surveys on HST is the ideal instrument to resolve such
components. Furthermore, to understand the nature of the aligned
blending, it will be very useful to obtain spectra from the ground. If
the aligned source is from a binary source companion, then
spectroscopy will reveal period shifts in the radial velocity. If the
light is from the lens, then cross-correlations of the spectra may
still show evidence for the lens due to the relative difference
between the radial velocities of the source and the lens (Mao, Reetz
\& Lennon 1998). These observations may be particularly valuable for
further understanding the lensed system, including a potential
determination of the lens mass.

It would be interesting to know how frequently these multiple peak
events occur, if indeed they are caused by the parallax
effect. Since an especially low velocity is required to produce such
multiple peaks, one can gain an indication of their prevalence by
studying theoretical distributions of the projected velocity,
$\tilde{v}$ (e.g., Han \& Gould 1995). The distributions
produced by Han and Gould (1995) suggest that it is extremely unlikely
such events could be caused by bulge-bulge lensing, since in this
instance values of $\tilde{v} < 100 \kms$ are strongly
disfavoured. For lenses residing in the disk, on the
other hand, values of $\tilde{v}$ are expected to be much smaller,
although velocities comparable to \sc40 also appear to be highly unlikely
according to their findings. One would need to perform detailed Monte
Carlo simulations to gain a firmer understanding of the prevalence of
such multiple peak events, and this is something which we are
currently undertaking. This is important because such events can easily
be overlooked when compiling microlensing catalogues since their
variability differs significantly from the standard microlensing light
curve and also because they take an unusually long time to reach a
constant baseline (due to the exceptionally small value of $\tilde{v}$
that is required to produce such multiple peaks). For example, the
variability for \sc40 is expected to last for well-over 5 years in
total. Therefore, if the observations had started once the event was
underway, this event could have easily been mistaken for a variable
source.

So far at least three convincing parallax candidates have been found
among the 520 microlensing events in the catalogue of Wo\'zniak et
al. (2001). These include \sc40, along with two events that have been
studied in previous papers, namely sc33\_4505 (Smith et al. 2002) and
OGLE-1999-BUL-32 (Mao et al. 2002). So it appears that the parallax
rate is about 1\%. However, as discussed in Smith et al. (2002), there
are a number of marginal microlensing events that were
uncovered. Detailed Monte Carlo simulations are needed to check
whether a similar rate is expected in the OGLE experiments. We are
currently performing such a study, the results of which will be reported
elsewhere.

\section*{Acknowledgement}

We thank Bohdan Paczy\'nski for discussions and encouragements.
We are deeply indebted to Andy Gould for a prompt and
comprehensive report which we implemented in \S5.
MCS acknowledges the financial support of a PPARC studentship.
PW was supported by the Laboratory Directed Research \& Development
funds (X1EM program at LANL). 
This work was also supported by the Polish grant KBN 2P03D01418.

{}

\begin{appendix}
\section{Rotating Binary-Source Model}
A comprehensive description of the binary-source motion should
consider elliptical orbits. However, since an exhaustive search of the
parameter space would prove to be demanding computationally, this
model is first simplified by considering only circular
orbits. The generalization to elliptical orbits is discussed in
\S\ref{app:elliptical}. A complete formalism can also be found in 
Dominik (1998).

\subsection{Circular orbits}
\label{app:circular}
The description of this circular-orbit binary-source model requires a
total of 12 parameters, i.e., a further 7 parameters in addition to
the 5 from the standard model. The motion of the two sources are
described using the following four parameters: the period of the
orbit, $T$; the binary source separation, $p$, which is given in units of the
Einstein radius projected into the source plane; the flux ratio of the
first source, ${\cal F}$, i.e.,
${\cal F}=\frac{f^{(1)}}{f^{(1)} + f^{(2)}}$ where $f^{(1)}$ and
$f^{(2)}$ are the fluxes from the first and second source respectively;
and the mass fraction of the first source, ${\cal M}$, i.e.,
${\cal M}=\frac{m^{(1)}}{m^{(1)} + m^{(2)}}$ where $m^{(1)}$ and
$m^{(2)}$ are the masses of the first and second source respectively.
In this prescription the
$x$-axis is defined such that source 1 lies on the (positive) $x$-axis at
$t=t_0$. There is no need to include a parameter to account for the
orbital phase because the orbits are circular and therefore rotationally
invariant. Unless otherwise stated, all of the distances in this section are
described in the orbital plane and given in units of the Einstein
radius projected into the source plane.

The above parameters allow the motion of the binary sources to be
determined, and the position vectors for the two sources,
${\mathbf{r}}^{(1)}$ and ${\mathbf{r}}^{(2)}$, are given by (see, for
example, Landau \& Lifshitz 1969), 
\beq \label{ap:r}
{\mathbf{r}}^{(i)}(t)=(\delta_{1,i}-{\cal M}){\mathbf{r}}(t), \;\;\;
{\rm where} \;\; {\mathbf{r}}(t)={p\:{\rm{cos}}(\xi) \choose
p\:{\rm{sin}}(\xi)} \;\;
{\rm and} \;\; \xi=\frac{2\pi}{T}(t-t_0), ~~ i=1, 2
\eeq
In this equation, $\delta_{1,i}$ is the usual Kronecker delta function,
and ${\cal M}$ is the mass ratio of the first source, as defined above.

As well as these parameters, a further three angles are required: two,
$\alpha$ and $\beta$, to determine the orientation of the orbital plane
(a longitude in the orbital plane, $\alpha$, and a latitude measured out
of the orbital plane, $\beta$, describing the line-of-sight vector to
the observer); and an angle to describe the lens trajectory in the
orbital plane, $\Theta$, which is measured from the $x$-axis towards the
direction of motion of the lens. When the above parameters
are added to the five from the standard model ($t_0$, $t_E$, $u_0$,
$f_{\rm s}$, $\Delta f$), this gives the complete set of parameters. However,
$u_0$ now corresponds to the minimum separation of the lens 
trajectory from the center of mass of the source system. This
separation is defined in the orbital plane, and is measured in units
of the projected Einstein radius. As with the parallax model, a
positive value of $u_0$ is chosen to correspond to the centre of mass
lying on the left-hand side of the lens trajectory. Also, $f_{\rm s}$ now
corresponds to the combined baseline flux of the two lensed sources
(i.e., $f_{\rm s}=f^{(1)} + f^{(2)}$).

This method is similar to that employed by Dominik (1998) in his study
of rotating binary sources. However, unlike Dominik's method, the
lens motion is defined in the orbital plane, rather than the lens
plane, and at this point only circular orbits are considered (see
above).

The lens position in the orbital plane is given by,
\beq \label{ap:rl}
{\mathbf{r}}_l(t)=u_0{{\rm{sin}}\Theta \choose -{\rm{cos}}\Theta}
+r_{{\rm{E}},p}(\Theta)\tau(t){{\rm{cos}}\Theta \choose {\rm{sin}}\Theta},
\eeq
\begin{displaymath}
{\rm{where}} \;\; \tau(t)=\frac{t-t_0}{t_E} \;\; {\rm{and}} \;\;
r_{{\rm{E}},p}(\Theta) = \frac{1}
{ \sqrt{ 1-{\rm{cos}}^2(\beta){\rm{cos}}^2(\Theta-\alpha) } }.
\end{displaymath}
This is analogous to the description of the lens trajectory given for
the parallax model in Soszy\'nski et al. (2001) (c.f. eq. 16 in
their paper).

Once these quantities have been determined then the next step is to
determine the separation between the source and the lens for both
sources individually, $\delta {\mathbf{r}}^{(i)} (t)$. This is simply
given by the vector from the lens position to the source position, i.e.,
$\delta {\mathbf{r}}^{(i)} (t) = {\mathbf{r}}^{(i)}(t) -
{\mathbf{r}}_l(t)$, where ${\mathbf{r}}^{(i)}(t)$ is the position of
source $i$ and ${\mathbf{r}}_l(t)$ is the position of the lens, as
defined in equations (\ref{ap:r}) and (\ref{ap:rl}), respectively.
Since $\delta {\mathbf{r}}^{(i)} (t)$ corresponds to the separation in
the orbital plane, this vector needs to be projected into the source
plane to obtain ${\mathbf{r}}^{(i)}_{\bot} (t)$, the component of this
separation perpendicular to the line-of-sight. This requires the
line-of-sight vector to the observer, ${\mathbf{\hat{n}}} =
({\rm{cos}}\beta\: {\rm{cos}}\alpha,\: {\rm{cos}}\beta\: {\rm{sin}}\alpha,\:
{\rm{sin}}\beta)$, and is given by,
\beq
\delta {\mathbf{r}}^{(i)}_{\bot} (t) = \delta {\mathbf{r}}^{(i)} (t)
- (\delta {\mathbf{r}}^{(i)} (t) {\mathbf{\cdot}}
{\mathbf{\hat{n}}}){\mathbf{\hat{n}}},
\eeq
Since all of the above distances are given in terms of the projected
Einstein radius, this quantity ${\mathbf{r}}^{(i)}_{\bot} (t)$ is
analogous to the separation parameter $u$ from the standard and
parallax models (see \S3).
The magnification of each source, $A^{(i)}(t)$, can then be calculated
independently by substituting $u^{(i)}(t) = \left| \delta
{\mathbf{r}}^{(i)}_{\bot} (t) \right|$ into the usual microlensing
equation (eq. \ref{amp}). This allows the lensed flux from each
source to be calculated individually, using the equation which was
employed in the previous models
(eq. \ref{eq:f}). These two fluxes can then be added together to give the
total flux. This addition of the two fluxes is justified since both
sources are able to be treated independently.
Therefore, the total flux is given by, 
\beq
f(t)=\left[ \left( A^{(1)}(t) - 1 \right){\cal F} + \left( A^{(2)}(t) - 1
\right)(1-{\cal F}) \right] f_{\rm s} + \Delta f,
\eeq
where $f_{\rm s}$ is the baseline flux of the lensed sources and $\Delta f$ is
the difference between the total baseline flux and the flux of the
reference image (this parameter, $\Delta f$, was introduced in
eq. \ref{eq:f}). This completes the prescription for the binary-source
model. However, it must be noted that a more complete model should
also take into account the parallax motion of the observer in addition
to the binary nature of the source.
In the case of LMC self-lensing events (e.g., Alcock et al. 2001b)
this effect should be negligible (see Gould 1998), but for \sc40 one
would expect this effect to be significant. This is because for \sc40
the source is closer to us (most probably located in the Galactic
bulge), hence the Einstein radius projected into the observer plane is
smaller, and so the parallax effect is expected to be larger (see \S3).

\subsection{Elliptical orbits}
\label{app:elliptical}
When considering elliptical orbits, two additional parameters are
required: the eccentricity, $e$; and the phase of the orbit,
$\xi_0$, which is defined as the angle between the $x$-axis and the
position of the first source at $t=t_0$. Also, since the orbits are
elliptical, the parameter $p$ now describes the orbital semi-major
axis, in units of the Einstein radius projected into the source
plane. 

The only difference between elliptical and circular orbits
comes into eq. (\ref{ap:r}), which describes the trajectory of the
binary sources. This equation should be replaced by its elliptical
counterpart,
\beq \label{ap:rell}
{\mathbf{r}}^{(i)}(t)=(\delta_{1,i}-{\cal M}){\mathbf{r}}(t), \;\;\;
{\rm where} \;\; {\mathbf{r}}(t)=
{p\:{\rm{cos}}(\xi) - e \choose
p\:\sqrt{1-e^2}\:{\rm{sin}}(\xi)}.
\eeq
The parameterization of this equation is now more complicated, with
$\xi$ being given by,
\beq
\xi - e\:{\rm{sin}}(\xi) =
2\pi \left( \frac{t-t_0}{T} \right) + \xi_0 - e\:{\rm{sin}}(\xi_0).
\eeq
Except for $e$ and $\xi_0$ (which were defined above), all of the
terms in this equation are as described in \S\ref{app:circular}. The
rest of the analysis is identical to that for circular orbits
(\S\ref{app:circular}).

\end{appendix}

\clearpage

\end{document}